\documentclass[conference]{IEEEtran}
\IEEEoverridecommandlockouts
\usepackage{cite}
\usepackage{amsmath,amssymb,amsfonts}
\usepackage{dutchcal}
\usepackage{fontawesome}
\usepackage[linesnumbered,ruled]{algorithm2e}
\usepackage{graphicx}
\usepackage{multirow}
\usepackage{subcaption} %
\usepackage{textcomp}
\usepackage[dvipsnames,table]{xcolor}
\usepackage{url}
\usepackage{listings}
\usepackage{xargs}
\usepackage{booktabs}
\usepackage{makecell}
\usepackage[colorinlistoftodos, prependcaption]{todonotes}
\usepackage[hidelinks, colorlinks=false, urlcolor=blue]{hyperref}
\usepackage{cprotect}
\usepackage{circledsteps}
\usepackage{annotate-equations}
\def\BibTeX{{\rm B\kern-.05em{\sc i\kern-.025em b}\kern-.08em
T\kern-.1667em\lower.7ex\hbox{E}\kern-.125emX}}
\usepackage{caption}
\SetCommentSty{mycommfont}
\usepackage{xparse}
\usepackage[]{mdframed}
\usepackage{enumitem}
\usepackage{doi}
\definecolor{lightgrayrow}{HTML}{EFEFEF}

\tikzset{annotate equations/text/.style={font=\bfseries\footnotesize}}
\NewDocumentCommand{\code}{v}{%
  \texttt{\textcolor{black}{#1}}%
}

\definecolor{ObservationBg}{HTML}{c7e9c0}

\newcounter{observationcounter}

\definecolor{circBG}{RGB}{186,90,84} %

\DeclareRobustCommand{\circled}[1]{%
  \tikz[baseline=(char.base)]{
    \node[circle, fill=circBG, inner sep=0pt] (char)
    {\color{white}\strut #1};
  }%
}

\makeatletter
\newcommand{\linebreakand}{%
\end{@IEEEauthorhalign}
\hfill\mbox{}\par
\mbox{}\hfill
\begin{@IEEEauthorhalign}
}
\makeatother

\begin{document}

\title{Every $\mu s$ Matters: Achieving Near Speed-of-Light Latency in GPU Collectives}

\author{
  \IEEEauthorblockN{Siyuan Shen\textsuperscript{*}\thanks{\textsuperscript{*}The majority of this work was done during an internship at NVIDIA.}}
  \IEEEauthorblockA{
    \textit{ETH Z\"urich}\\
    Z\"urich, Switzerland \\
  siyuan.shen@inf.ethz.ch}
  \and
  \IEEEauthorblockN{Anton Korzh}
  \IEEEauthorblockA{
    \textit{NVIDIA Corporation}\\
    Santa Clara, California \\
  akorzh@nvidia.com}
  \and
  \IEEEauthorblockN{John Bachan}
  \IEEEauthorblockA{
    \textit{NVIDIA Corporation}\\
    Santa Clara, California \\
  jbachan@nvidia.com}
  \and
  \IEEEauthorblockN{Tiancheng Chen}
  \IEEEauthorblockA{
    \textit{ETH Z\"urich}\\
    Z\"urich, Switzerland \\
  tiancheng.chen@inf.ethz.ch}
  \and
  \IEEEauthorblockN{Arnav Goel}
  \IEEEauthorblockA{
    \textit{NVIDIA Corporation}\\
    Santa Clara, California \\
  arnavg@nvidia.com}
  \and
  \IEEEauthorblockN{Ludwig Schneider}
  \IEEEauthorblockA{
    \textit{NVIDIA Corporation}\\
    Santa Clara, California \\
  lschneider@nvidia.com}
  \and
  \IEEEauthorblockN{Pouya Kousha}
  \IEEEauthorblockA{
    \textit{NVIDIA Corporation}\\
    Santa Clara, California \\
  pkousha@nvidia.com}
  \and
  \IEEEauthorblockN{Zhenhao He}
  \IEEEauthorblockA{
    \textit{NVIDIA Corporation}\\
    Z\"urich, Switzerland \\
  zhenhaoh@nvidia.com}
  \and
  \IEEEauthorblockN{Sylvain Jeaugey}
  \IEEEauthorblockA{
    \textit{NVIDIA Corporation}\\
    Grenoble, France \\
  sjeaugey@nvidia.com}
  \linebreakand
  \IEEEauthorblockN{Kamil Iskra}
  \IEEEauthorblockA{
    \textit{NVIDIA Corporation}\\
    Chicago, Illinois \\
  kiskra@nvidia.com}
  \and
  \IEEEauthorblockN{Nishank Chandawala}
  \IEEEauthorblockA{
    \textit{NVIDIA Corporation}\\
    Santa Clara, California \\
  nchandawala@nvidia.com}
  \and
  \IEEEauthorblockN{Jeff R. Hammond}
  \IEEEauthorblockA{
    \textit{NVIDIA Helsinki Oy}\\
    Helsinki, Finland \\
  jeffpapers@nvidia.com}
  \and
  \IEEEauthorblockN{Torsten Hoefler}
  \IEEEauthorblockA{
    \textit{ETH Z\"urich}\\
    Z\"urich, Switzerland \\
  torsten.hoefler@inf.ethz.ch}
}

\maketitle
\IEEEpeerreviewmaketitle
\thispagestyle{plain}
\pagestyle{plain}

\begin{abstract}
GPU collective communication is typically optimized for bandwidth, yet many emerging workloads are increasingly limited by latency. Long-context decode-heavy large language model (LLM) inference is a prime example, where serving large models requires multiple GPUs, and many small collectives lie directly on the critical path of token generation. Therefore, even $\mu s$ of overhead can impact performance and cost. In this work, we study how to approach the hardware Speed-of-Light (SoL) lower bound for GPU collectives within a scale-up network. We identify key principles for near-optimal designs, including barrier-free synchronization and efficient use of symmetric memory and multicast. Building on NCCL's device-side API, we develop low-latency interfaces for constructing custom collective kernels and use them to implement new symmetric collectives in NCCL. Microbenchmarks show substantial latency reductions for small and medium messages, reducing overhead to within 7\% of the absolute SoL lower bound. When integrated into real applications, these kernels improve inter-token latency and throughput in LLM inference and accelerate cuSOLVERMp, demonstrating benefits for both AI inference and traditional HPC workloads.
\end{abstract}

\section{Introduction}

\begin{figure}[!t]
  \centering
  \includegraphics[width=.95\linewidth]{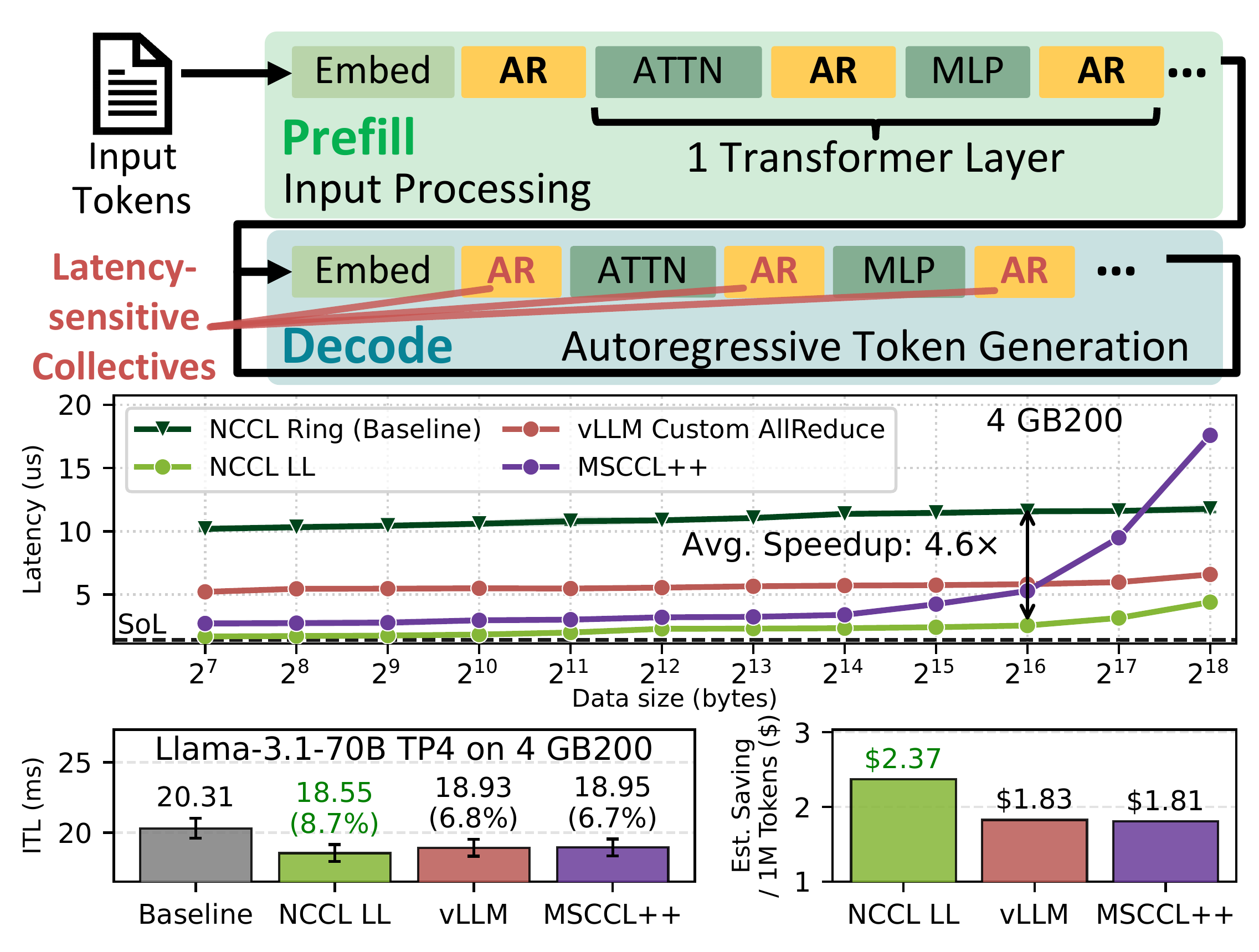}
  \cprotect\caption{In long-context, small-batch tensor-parallel (TP) LLM inference, many small AllReduce operations lie on the critical path, making collective latency a crucial bottleneck. The microbenchmark shows that our NCCL low-latency kernel reduces small-message AllReduce latency relative to other implementations, approaching the speed-of-light (SoL) bound. This translates into lower inter-token latency (ITL) and higher cost savings for Llama-3.1-70B inference.}
  \label{fig:poster-child}
\end{figure}

Deep learning has driven rapid growth in GPU cluster scale and interconnect capability. As large language models (LLMs) scale, inference increasingly spans multiple GPUs, bringing collective communication onto the critical path of token generation. For example, serving DeepSeek-V3-class models~\cite{deepseekai2025deepseekv3} requires at least 8$\times$H200 GPUs~\cite{deepseek_v3_hw_req_alt}. The impact of collective communication is especially evident in decode-heavy workloads, where modern systems may generate millions of tokens per request in applications such as code generation and agent-style workflows~\cite{kwon2023vllm, aubakirova2026stateaiempirical100}. In these settings, even small communication overheads accumulate and directly impact the quality of service. In addition, for long-context inference, the KV-cache memory grows with sequence length, and the batch size is often reduced to fit within device memory. As a result, collectives such as AllReduce are invoked frequently with relatively small message sizes during decoding, making latency, rather than bandwidth, the dominant bottleneck. Consequently, recent inference frameworks such as vLLM~\cite{kwon2023vllm}, SGLang~\cite{zheng2024sglang}, and TensorRT-LLM~\cite{nvidia2023trtllm} treat communication latency as a first-class optimization target and implement custom low-latency GPU kernels for operations such as AllReduce.

Low-latency collectives are essential not only for LLM inference, but also for many traditional scientific and HPC applications. Many simulations and solvers perform frequent small global reductions within tightly synchronized phases, such as time-stepping loops and particle simulations. In these settings, collective latency lies on the critical path and can limit strong scaling. Prior work such as LLAMP shows that widely used HPC workloads, including MILC and LULESH, are measurably sensitive to collective latency~\cite{shen2024llamp}. At the same time, many GPU-accelerated scientific applications do not yet fully exploit GPU-native communication libraries.
These observations indicate that low-latency GPU collectives can improve not only LLM inference, but also scalability and time-to-solution in traditional scientific workloads.

Despite this growing recognition, existing approaches still leave performance on the table. We observed that even the best available implementations often remain above the \textbf{“speed-of-light”} (SoL) bound, by which we mean the absolute hardware lower bound imposed by the interconnect and memory system. Figure~\ref{fig:poster-child} illustrates this effect for the long-context, small-batch decode setting that we target. On 4 GB200 GPUs, the NCCL low-latency kernel we introduce in this work reduces average latency for small messages from 11.0~$\mu$s for NCCL ring to 2.37~$\mu$s, yielding an 8.7\% ITL reduction for the inference workload of Llama-3.1-70B. Using CoreWeave's on-demand price of \$42/hour for 4 GB200~\cite{coreweave2026pricing} and converting output throughput into cost per 1M output tokens, the measured data implies that each $\mu s$ removed from AllReduce latency reduces cost by about 0.9\%. While this saving may appear insignificant, it compounds into substantial cost reduction at the trillion-token scale of modern LLM services~\cite{aubakirova2026stateaiempirical100,microsoft2022billionsinferences}.

In this work, we begin by identifying global memory barriers across participating GPUs as a key source of latency in existing collective implementations. To this end, we present several techniques, including LL, sentinel-based synchronization, double buffering, and a novel two-shot AllReduce algorithm. By combining these techniques, we eliminate expensive global memory barriers entirely while preserving correctness and efficiency. Building on these, we develop a set of experimental application programming interfaces (APIs) on top of NCCL’s latest device communication APIs that encapsulate these low-latency mechanisms into reusable primitives for efficiently prototyping custom low-latency kernels. Leveraging this interface, we implement several new AllReduce kernels within NCCL that are directly usable in practice.

To evaluate these techniques and the resulting collectives, we conduct detailed microbenchmarks showing that our designs approach the hardware SoL latency bound across a wide range of node configurations. We also integrate our low-latency kernels into real workloads, including vLLM and cuSOLVERMp. These case studies demonstrate consistent and measurable performance improvements over standard NCCL collectives and other state-of-the-art frameworks, confirming the practical benefits of latency-centric collective optimization for both LLM inference and traditional HPC workloads.

Our contributions in this work are as follows:
\begin{itemize}[leftmargin=1em]
  \item We systematically analyze existing techniques for reducing collective latency, characterize where they are effective, and compose them into barrier-free designs that approach the hardware lower bound.
  \item We design and implement a set of low-latency communication APIs on top of NCCL’s device-side APIs to facilitate development of custom collective kernels.
  \item We develop new low-latency AllReduce algorithms, including a novel two-shot LL128 atomic design, using the proposed low-latency APIs.
  \item We conduct extensive evaluations through microbenchmarks and real workloads, including vLLM inference and cuSOLVERMp, demonstrating noticeable performance gains.
\end{itemize}

\section{Background}

\subsection{GPU Communication Libraries}

Efficient communication is essential in modern GPU-accelerated systems, where both AI and scientific workloads rely on tightly coupled GPU execution~\cite{fusco2024understanding}. To support this, specialized GPU communication libraries have emerged as a critical software layer. NCCL is one of the most widely used libraries, providing collective and point-to-point operations optimized for various interconnects~\cite{hu2025demystifying}. While NCCL primarily targets optimized collectives, NVSHMEM adopts a partitioned global address space (PGAS) model that enables one-sided communication and exposes finer device-side control~\cite{nvidia2022nvshmem}. Similar libraries exist across vendors, including AMD’s RCCL and rocSHMEM, and Intel’s oneCCL~\cite{rccl2024,amd2026rocshmem,oneccl2024}. In contrast, traditional frameworks like MPI~\cite{mpi50}, originally designed for CPU-based systems, have been extended to support GPUs (e.g., CUDA-aware MPI~\cite{kraus2025cudaawarempi}), but often underperform vendor-optimized libraries for large-scale collectives while remaining competitive for point-to-point communication~\cite{desensi2024exploring}. These observations highlight that GPU-native communication libraries have become an essential complement to traditional frameworks in modern HPC and AI systems.

\begin{figure}[!t]
  \centering
  \includegraphics[width=1\linewidth]{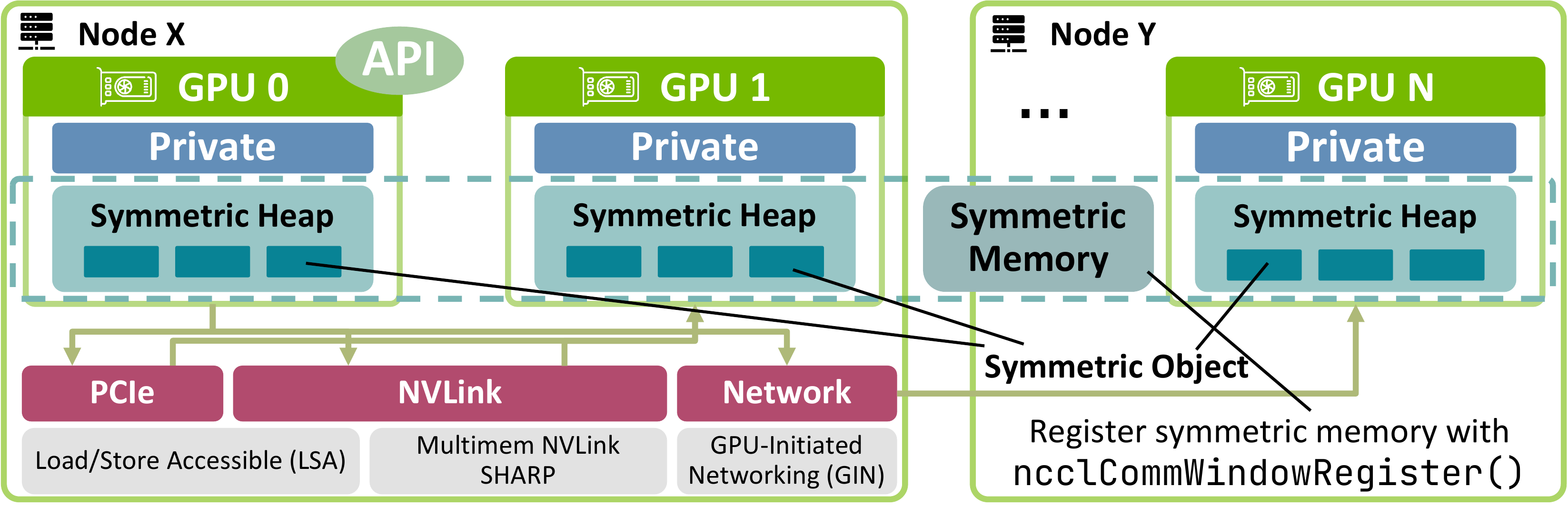}
  \cprotect\caption{Overview of device-initiated communication and symmetric memory in NCCL~\cite{bachan2025enabling}. When GPUs are in the same node or NVLink domain, LSA operations are supported over PCIe and NVLink, while multimem operations are enabled via NVLink SHARP for hardware-accelerated multicast and reduction. For inter-node communication, GPU-initiated networking (GIN) supports GDAKI and proxy-assisted data transfers over InfiniBand and RoCE~\cite{hamidouche2025gin}.}
  \label{fig:background-sym-mem}
\end{figure}

\subsection{Device-Initiated Communication and Symmetric Memory}

In addition to optimized collective primitives, modern CCLs increasingly support device-driven communication, enabling kernels to directly initiate and orchestrate data movement. Early support for this appeared in NVSHMEM, which allows kernels to perform remote memory operations (e.g., put/get) and synchronization without host involvement. More recently, NCCL 2.28 introduces device-side communication APIs that enable kernels to directly invoke communication primitives~\cite{nvidia_nccl_deviceapi, hamidouche2025gin}. These capabilities are enabled by underlying hardware and runtime support, including GPU Virtual Memory Management (VMM), which provides a unified virtual address space across GPUs, and GPUDirect Async Kernel-Initiated (GDAKI), which allows GPUs to directly interact with network interfaces without CPU intervention.

Symmetric memory originates from the SHMEM family of PGAS models. Remotely accessible data objects, called \textbf{symmetric objects}, have identical type, size, and layout on each processing element (PE), which corresponds to a GPU in this case. This allows remote access using the same logical address together with a PE identifier. These objects reside in the \textbf{symmetric heap}, which supports one-sided operations such as get, put, and atomics.

On GPUs connected through PCIe or NVLink and supported by CUDA Virtual Memory Management (VMM), symmetric memory regions can be mapped into a unified virtual address space, making them \textbf{load/store accessible (LSA)}. On systems with NVSwitch and NVLink SHARP (NVLS), \textbf{multimem load/store} instructions can further accelerate communication by enabling multicast and in-network reduction. A visualization is shown in Fig.~\ref{fig:background-sym-mem}. Overall, symmetric memory reduces address translation overhead and enables low-latency data exchange. Thus, NCCL is gradually replacing its collectives with symmetric-memory-based kernels, and previous implementations are now referred to as \textbf{legacy kernels}~\cite{jeaugey2025nccl228, bachan2025enabling}.

We base our implementation on NVIDIA hardware and NCCL because this stack represents one of the most widely adopted platforms for GPU collectives in modern AI and HPC ecosystems~\cite{pytorch_distributed, nhrfau_pytorch}. We chose NCCL rather than NVSHMEM since NCCL is extensively used as a communication backend across both deep learning frameworks and other scientific libraries, including PyTorch~\cite{paszke2019pytorch}, TensorFlow~\cite{abadi2015tensorflow}, vLLM~\cite{kwon2023vllm}, cuSOLVERMp~\cite{nvidia_cusolver_api}, and cuBLASMp~\cite{nvidia_cublasmp}. This level of integration makes NCCL a more suitable choice.

\textbf{Most of the design principles are, nevertheless, not NVIDIA-specific.} LL, sentinel synchronization, and double buffering require GPU-initiated access to peer memory, remote writes that become visible to GPU-side polling, and device-side ordering or fence operations before buffer reuse. All platforms providing these properties can implement the proposed protocols and kernels.

Additionally, this work focuses on collectives within a scale-up network, specifically GPUs residing in the \textbf{same NVLink domain}. We exclude scale-out communication for several reasons. First, in modern LLM inference, which is a primary target of this work, parallel groups are typically confined to a single scale-up domain. Second, multi-node systems commonly employ hierarchical collectives that separate local and scale-out phases, making improvements within the scale-up domain complementary to higher-level optimizations~\cite{singhania2025llminferencesinglenode, hwang2026msccl}. Finally, emerging GPU systems increasingly expand the size and capability of these domains, allowing a growing fraction of latency-sensitive workloads to execute entirely within a single scale-up network~\cite{patel2025gtc, xai2024colossus}.

\section{Toward Near Speed-of-Light AllReduce}

In this section, we describe how we approach near speed-of-light (SoL) latency, i.e., the hardware lower bound, for \textbf{AllReduce} (R is capitalized following NCCL's notation). We focus on AllReduce because it is one of the most widely used collectives in HPC and distributed machine learning and a frequent optimization target in practice~\cite{tang2024recent, xiong2024revisiting, adam2023xccl, chrapek2023hear, bernholdt2018asurvey, laguna2019alargescale}. Moreover, many implementations decompose AllReduce into ReduceScatter and AllGather or Reduce and Broadcast, so techniques that minimize AllReduce latency often apply directly to these building blocks. Thus, optimizing AllReduce benefits a broader class of collectives.

\subsection{Low-Latency AllReduce Algorithms}
\label{sec:low-latency-ar}

When the message size of an AllReduce operation is small, \textbf{its latency is primarily determined by the number of synchronizations needed}, or communication phases. Consequently, algorithms such as tree-based or recursive-doubling AllReduce, which require $O(\log N)$ rounds of synchronization for $N$ ranks, typically outperform ring-based algorithms that require $O(N)$ rounds for small messages~\cite{hu2025demystifying, hammer2025shortcircuiting}. In a scale-up network, the number of synchronizations can be reduced further to $O(1)$ using one-shot or two-shot AllReduce algorithms, which are widely adopted in most communication libraries and frameworks~\cite{kwon2023vllm, hu2025demystifying, nvidia2023trtllm, nvidia2022nvshmem, hwang2026msccl}.

\subsubsection{One-shot AllReduce}

In a one-shot AllReduce, the entire reduction is completed in a single communication phase, where each GPU fetches data from all peers, performs the reduction locally, and writes the result to the output. In \textbf{pull} mode, GPUs read remote data via loads, while in \textbf{push} mode they write data to remote buffers before reducing locally. Push is generally faster, requiring only half a GPU-to-GPU RTT versus a full RTT for remote loads, but needs additional buffering for incoming data. Consequently, push-based one-shot designs are often preferred for latency-sensitive collectives.

\subsubsection{Two-shot AllReduce}

\begin{figure}[!t]
  \centering
  \includegraphics[width=1\linewidth]{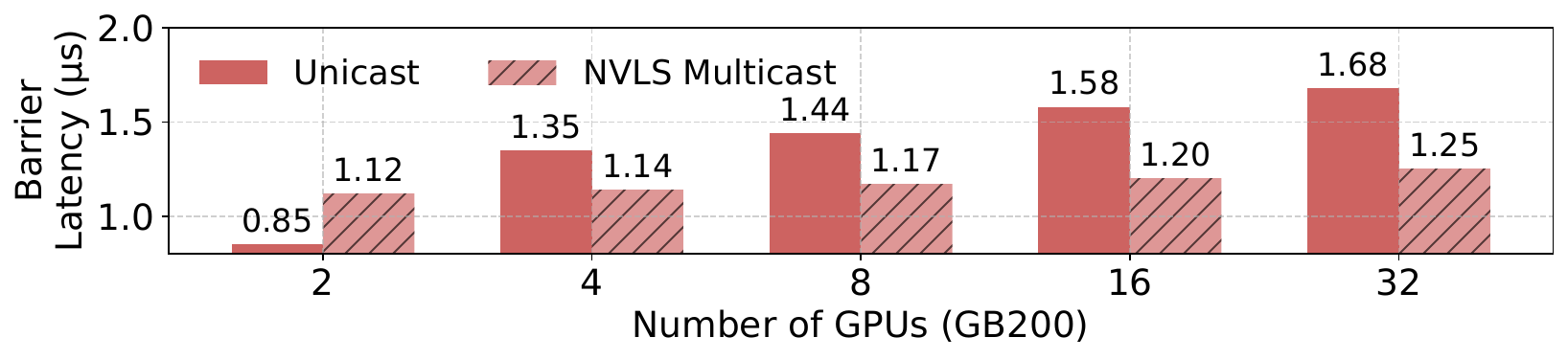}
  \cprotect\caption{Barrier latency on GB200 as a function of the number of GPUs for unicast and multicast implementations.}
  \label{fig:barrier-latency}
\end{figure}

\begin{figure*}[!h]
  \centering
  \includegraphics[width=.85\linewidth]{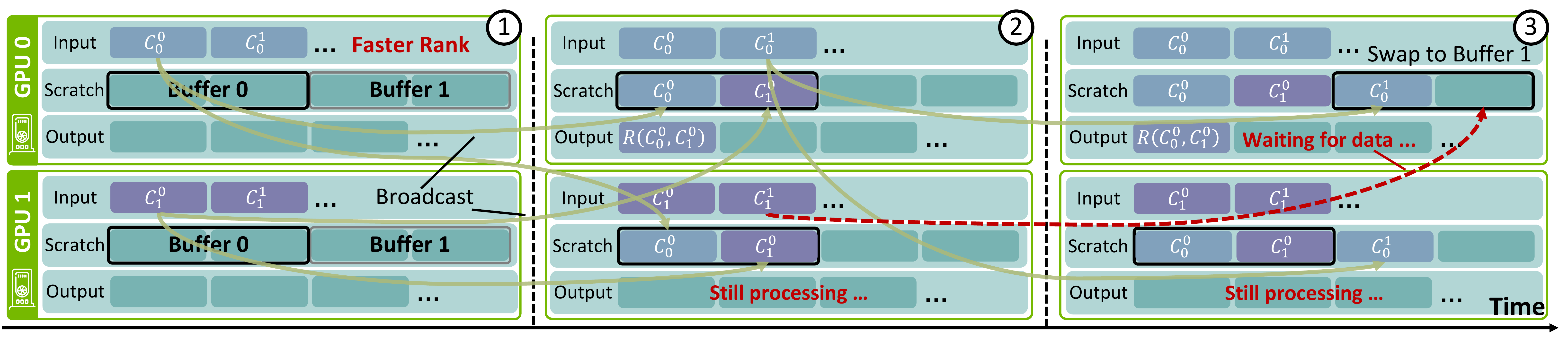}
  \cprotect\caption{Example illustrating bidirectional communication with double buffering. Two ranks exchange chunked inputs, where chunk $i$ from rank $r$ is denoted $C_r^i$. The scratch space is divided into Buffers 0 and 1, each can store two chunks. The black border marks the active buffer. Steps \textcircled{1}, \textcircled{2}, and \textcircled{3} show events in chronological order, with time progressing to the right. Arrows represent cross-GPU data transfer. The mechanism is independent of whether LL or sentinel synchronization is used.}
  \label{fig:double-buffering-example}
\end{figure*}

A two-shot AllReduce is decomposed into two phases. In the ReduceScatter phase, each GPU exchanges partitions of its input with peers and performs the reduction on its assigned chunk, producing a partial result. Either push or pull semantics may be used in this phase. In the AllGather phase, the reduced chunks are exchanged so that every GPU has the complete result. Compared to one-shot, two-shot introduces an additional synchronization but significantly reduces the communication volume, improving performance for moderate message sizes. For $N$ ranks reducing $M$ bytes of data, one-shot incurs $O(N\cdot M)$ total communication volume, whereas two-shot reduces this to $O(M)$.

\subsection{Cost of Memory Barriers}

After examining several one-shot and two-shot AllReduce implementations in state-of-the-art frameworks and libraries~\cite{nvidia2023trtllm, nvidia2022nvshmem, kwon2023vllm, hwang2026msccl}, we observe that, regardless of push or pull communication, these designs typically rely on explicit memory barriers to synchronize peers and signal data readiness at the thread-block level. Using NCCL’s \texttt{ncclLsaBarrierSession}, which implements a memory barrier for load-store accessible (LSA) devices, we measure the overhead of such synchronization under relaxed memory ordering and report the results in Fig.~\ref{fig:barrier-latency}. Although NCCL’s implementation may not be fully optimized, alternative designs follow the same fundamental pattern: a flag is propagated to all peers, and each GPU waits until it observes the corresponding signals from every other participant. Therefore, it is representative of the inherent cost in such approaches.

The measured barrier latency shows that each barrier will incur more than 1 $\mu s$ of overhead. In many AllReduce kernels, two such barriers are required. As illustrated in Fig.~\ref{fig:poster-child}, when a small-message AllReduce completes in roughly 5 $\mu s$ on four GPUs, two barrier calls alone will account for about 40\% of the total latency. As emphasized earlier, when targeting near SoL performance, every $\mu s$ matters. Thus, \textbf{eliminating these barriers entirely can yield substantial performance gains}.

\section{Designing Barrier-Free Collectives}

Having established that memory barriers introduce non-negligible overhead, we next consider alternative synchronization mechanisms that achieve the same purpose with lower latency. As discussed in Section~\ref{sec:low-latency-ar}, AllReduce can be implemented using either push or pull communication. Since our goal is to approach the SoL, we focus on \textbf{push mode}, which trades additional buffer space, referred to here as a \textbf{scratch buffer}, for roughly half of a GPU-to-GPU RTT.

\subsubsection{LL}

The first technique we introduce is LL, short for low latency, which originates from the LL protocol in NCCL~\cite{hu2025demystifying} and is also used in libraries such as NVSHMEM~\cite{nvidia2022nvshmem} and MSCCL++~\cite{hwang2026msccl}. Conventional synchronization signals data arrival using explicit flags and enforced ordering between data and signals. LL removes the signaling step by packing the 8-byte flag with the 8-byte data and transmitting them atomically with 16-byte atomic stores, allowing the receiver to determine data readiness by checking the flag directly. This design halves the effective payload bandwidth and doubles scratch buffer usage, making LL mostly suitable for very small messages.

\subsubsection{Sentinel}

Instead of embedding a signal in the transmitted data, the receiving scratch buffer can be initialized with a sentinel value that is unlikely to appear in valid computations, such as the floating-point value \texttt{-NaN}. Data is then written directly to the buffer, and the receiver polls until the value changes from the sentinel, indicating that valid data has arrived. Compared to LL, this approach preserves full effective bandwidth and uses less scratch space, making it more efficient for moderately larger messages. However, it has a few drawbacks. First, unlike LL, which updates its flag each iteration, the sentinel method requires the buffer to be reset before reuse, complicating buffer management. Second, transmitted values must never match the sentinel because a matching value would prevent the receiver from detecting data arrival. Users must therefore exclude such values to ensure correctness.

\subsubsection{Bidirectional Communication \& Double Buffering}

Although LL and sentinel synchronization eliminate explicit memory barriers for single exchanges, they are insufficient when messages require multiple iterations due to limited buffer space. In such cases, inputs are partitioned and processed in chunks. Conventional approaches insert barriers between iterations to prevent buffer overwrites. We can eliminate them by using bidirectional communication and double buffering.

Fig.~\ref{fig:double-buffering-example} demonstrates the mechanism. Consider two ranks whose scratch space cannot hold all input data simultaneously. In step \textcircled{1}, both ranks begin with the scratch buffer set to Buffer 0. In step \textcircled{2}, they exchange their first chunks $C_r^0$. Because Rank 0 processes data faster, it completes the reduction $R$ and writes the result to its output while Rank 1 is still processing. In step \textcircled{3}, Rank 0 advances to the next iteration and switches to Buffer 1 to broadcast chunk $C_0^1$. It then waits for $C_1^1$ before proceeding, ensuring that data in Buffer 0 is not overwritten before Rank 1 finishes reading it.

This bidirectional exchange, combined with double buffering, ensures that a rank cannot overwrite a peer’s buffer for the next iteration until it has received data from that peer in the current one. This assumes that each rank communicates with a given peer at most once per iteration, avoiding multiple stores to the same remote address. In effect, \emph{each receive from a peer serves as an implicit permission for the next send}, analogous to credit-based flow control. This mechanism allows multiple reduction iterations without costly global memory barriers.

\begin{figure}[!t]
  \centering
  \includegraphics[width=1\linewidth]{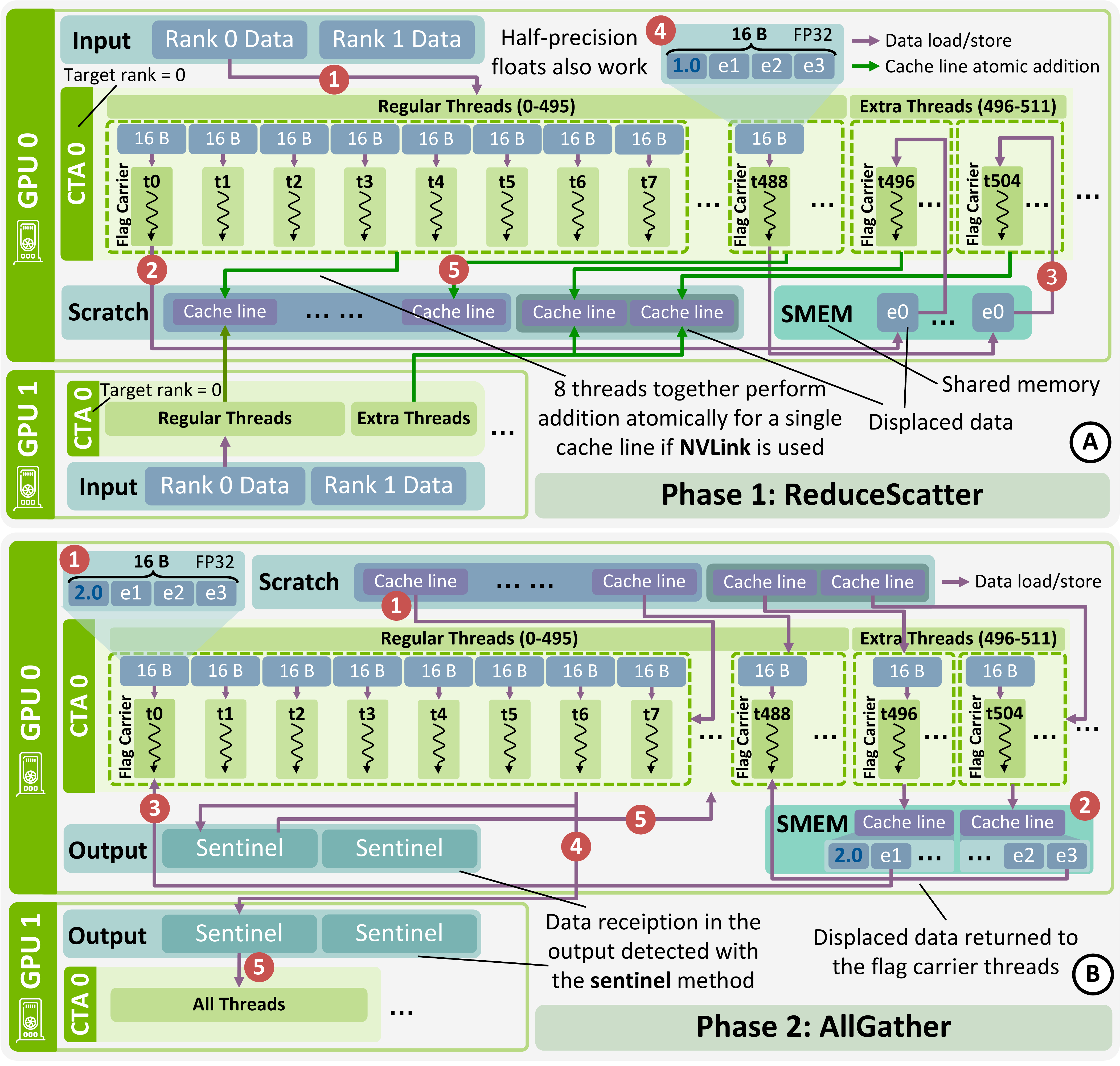}
  \cprotect\caption{Overview of the two-shot LL128 atomic AllReduce algorithm. Threads within a CTA are divided into two groups: 496 \textbf{regular threads} and 16 \textbf{extra threads}, which handle displaced elements. The dashed green boxes indicate groups of 8 threads that operate on a 128-byte cache line. Panel \textcircled{A} illustrates the ReduceScatter stage, while Panel \textcircled{B} shows the AllGather stage. The example assumes an out-of-place operation where the output buffer is initialized with sentinel values before kernel execution. To support half-precision floats, each element is only 2 bytes and only 8 extra threads are needed.}
  \label{fig:new-algo}
\end{figure}

\begin{table}[!t]
  \renewcommand{\arraystretch}{1.1}
  \resizebox{\columnwidth}{!}{%
    \begin{tabular}{ccccc}
      \toprule
      \textbf{
        \begin{tabular}[c]{@{}c@{}}AllReduce\\Algorithm
      \end{tabular}} &
      \textbf{
        \begin{tabular}[c]{@{}c@{}}Comm. Volume\\per GPU
      \end{tabular}} &
      \textbf{
        \begin{tabular}[c]{@{}c@{}}Latency,\\\# Synchronizations
      \end{tabular}} &
      \textbf{
        \begin{tabular}[c]{@{}c@{}}Scratch\\Space / Iter
      \end{tabular}} &
      \textbf{Deterministic} \\[2pt] \hline

      \rowcolor{lightgrayrow}
      \vrule height 4ex depth 2.5ex width 0pt
      \textbf{
        \begin{tabular}[c]{@{}c@{}}One-shot\\ (LL)
      \end{tabular}}               & $2 (N -1)  M$                  & 1 & $2ND$         & \faCheck \\

      \vrule height 4ex depth 2.5ex width 0pt
      \textbf{
        \begin{tabular}[c]{@{}c@{}}One-shot\\ (Sentinel)
      \end{tabular}}         & $(N -1) M$                     & 1 & $N D$          & \faCheck \\

      \rowcolor{lightgrayrow}
      \vrule height 4ex depth 2.5ex width 0pt
      \textbf{
        \begin{tabular}[c]{@{}c@{}}Two-shot\\ (LL)
      \end{tabular}}               & $4 (N - 1) \frac{M}{N}$         & 2 & $2D$          & \faCheck \\

      \vrule height 4ex depth 2.5ex width 0pt
      \textbf{
        \begin{tabular}[c]{@{}c@{}}Two-shot\\ (Sentinel)
      \end{tabular}}         & $2(N - 1) \frac{M}{N}$          & 2 & $D$           & \faCheck \\

      \rowcolor{lightgrayrow}
      \vrule height 4ex depth 2.5ex width 0pt
      \textbf{
        \begin{tabular}[c]{@{}c@{}}Two-shot\\ (LL128 Atomic)
      \end{tabular}}            & $\approx 2 (N - 1) \frac{M}{N}$ & 2 & $\approx\frac{D}{N}$ & \faClose \\

      \hline
    \end{tabular}%
  }
  \caption{Comparison of low-latency AllReduce algorithms. $N$ denotes the number of GPUs, $M$ the total message size, and $D$ the amount of data reduced per iteration. Latency is expressed as the number of synchronizations required per iteration. ``\emph{Scratch Space / Iter}'' denotes the scratch buffer capacity required to reduce $D$ bytes of data per iteration.}
  \label{tab:ar-algo-comparison}
\end{table}

\subsubsection{LL128 Atomic AllReduce}

The last technique we present for removing the global memory barrier is a new AllReduce algorithm that uses a synchronization mechanism different from LL and sentinel. Since it resembles NCCL’s LL128 protocol~\cite{hu2025demystifying} and relies on atomic additions, we refer to it as the \textit{two-shot LL128 atomic} algorithm. An overview is shown in Fig.~\ref{fig:new-algo}. Like the standard two-shot design, the algorithm proceeds in two phases, which we describe in detail below.

\paragraph{ReduceScatter} As in standard ReduceScatter algorithms, the input is partitioned into $N$ chunks, where $N$ is the number of GPUs. To exploit GPU parallelism, CTAs are evenly distributed across ranks, and each CTA is assigned a \textbf{target rank}. For example, in Fig.~\ref{fig:new-algo}, CTA 0 on both GPU 0 and 1 processes the chunk belonging to GPU 0. Each CTA then reads its assigned partition and performs the steps below.

\noindent\circled{1}\hspace{0.3em}
Threads operate in groups of 8, and each thread processes 16 bytes from the assigned partition. For FP32 data, each thread handles 4 elements ($e0$–$e3$). Together, the 8 threads operate on 128 bytes, which matches the size of a cache line.

\noindent\circled{2}\hspace{0.3em}
Within each group, the first thread acts as the \textbf{flag carrier}. It moves its first element $e_0$ into a shared memory region reserved for displaced values. A \texttt{\_\_syncthreads()} then ensures that the displaced data is visible to all threads.

\noindent\circled{3}\hspace{0.3em}
Threads in the extra-thread group read the displaced elements from shared memory.

\noindent\circled{4}\hspace{0.3em}
Each flag carrier sets the first element of its vector to 1.

\noindent\circled{5}\hspace{0.3em}
All threads then perform an \textbf{atomic add} to the scratch buffer corresponding to the target rank. NVLink ensures that these operations are applied atomically at the cache-line level.

\paragraph{AllGather}
Since NVLink performs 128-byte writes atomically, when the first element of a cache line (the flag) equals the number of ranks $N$, it indicates that all ranks have contributed. The AllGather phase then proceeds as follows.

\noindent\circled{1}\hspace{0.3em}
Each CTA whose target rank matches its own rank polls the corresponding region in the scratch buffer. Within each group of eight threads, the flag carrier repeatedly checks the first element of its vector until the value equals $N$.

\noindent\circled{2}\hspace{0.3em}
Once the data is confirmed to be ready, the extra threads write their 16-byte displaced elements into shared memory. A \texttt{\_\_syncthreads()} is executed afterwards to ensure visibility of data. 

\noindent\circled{3}\hspace{0.3em}
The flag carriers from the regular thread group then read the displaced elements from shared memory, restoring the correct element within their vectors.

\noindent\circled{4}\hspace{0.3em}
All regular threads write their data to the corresponding region of the output buffer.

\noindent\circled{5}\hspace{0.3em}
CTAs then poll the output buffer partitions associated with their target ranks until the complete data becomes available.

\subsubsection{Algorithm Comparison}

\begin{figure*}[!t]
  \centering
  \includegraphics[width=1\linewidth]{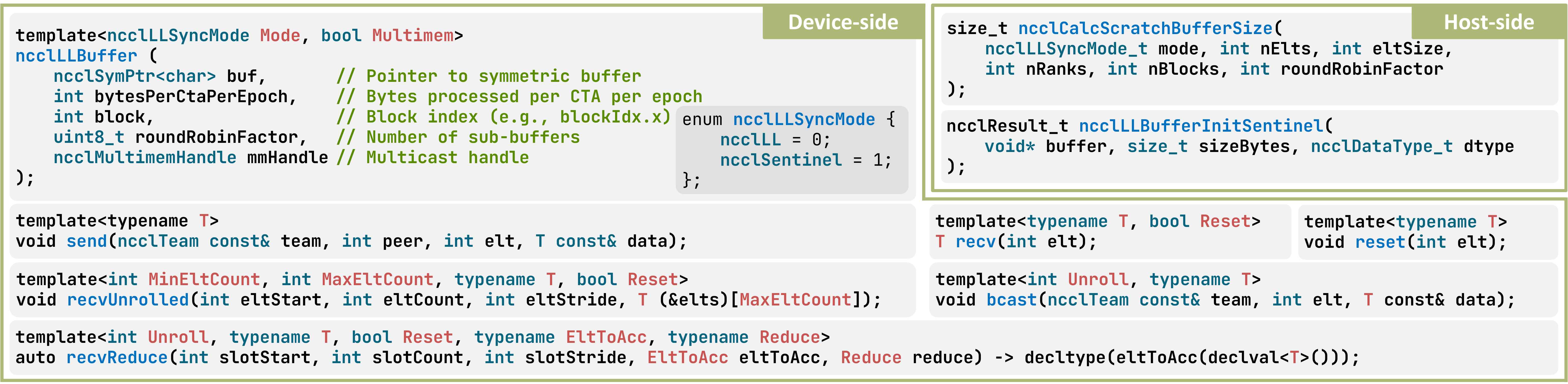}
  \cprotect\caption{Overview of the proposed low-latency API, showing device-side and host-side functions.}
  \label{fig:ll-apis}
\end{figure*}

The two-shot LL128 atomic algorithm was proposed to improve scalability. The motivation is twofold. First, atomic additions synchronized at the L2 cache may be cheaper than waiting for all data and performing reductions within CTAs. Second, the required scratch buffer space is only $\frac{D}{N}$. The algorithm also wastes far less bandwidth than LL. For 32-bit floats, it requires only 4 extra bytes per 128 bytes of data ($\approx 3\%$). For 16-bit floats, it requires 2 extra bytes per 128 bytes ($\approx 1.5\%$).

The algorithm has several limitations. It requires NVLink to guarantee cache-line–level atomic addition and supports only single- and half-precision types due to the availability of vectorized atomics in CUDA~\cite{nvidiacuda, nvidiaptx}. It is also limited to addition, as the embedded flag relies on commutativity, and CUDA does not provide vectorized atomic multiplication. In practice, this is not too restrictive since addition dominates most target workloads, such as LLM inference. Finally, the algorithm is non-deterministic because floating-point atomic ordering is not guaranteed. A one-shot variant is possible, but different ranks may observe different results, which violates AllReduce semantics~\cite{mpi50}. The trade-offs of different low-latency algorithms are summarized in Table~\ref{tab:ar-algo-comparison}.

In terms of numerical stability, the algorithm obeys the standard forward-error bound for floating-point summation. For $N$ ranks and unit roundoff $u$ in the accumulation format, assuming no overflow or underflow,
\[
  \left|\operatorname{fl}\!\left(\sum_{i=1}^{N}x_i\right)-\sum_{i=1}^{N}x_i\right|
  \leq \gamma_{N-1}\sum_{i=1}^{N}|x_i|,
  \;
  \gamma_k=\frac{ku}{1-ku}.
\]
As an example, for FP32 and 64 ranks, the worst-case coefficient is $\gamma_{63}\approx3.8\times10^{-6}$. The bound is larger for FP16 and BF16. Thus, LL128 atomic should be treated as a performance-oriented option for workloads that tolerate the precision of the selected accumulation format.

\section{Low-Latency API Design}

Building on the techniques described previously, we translate them into APIs designed to meet three requirements. \textcircled{1} Compatibility with NCCL’s existing device-side interface. \textcircled{2} Encapsulation of the low-latency techniques introduced earlier through a unified abstraction. \textcircled{3} Sufficient flexibility to serve as building blocks for custom communication kernels. Guided by these principles, we develop a set of experimental low-latency APIs and integrate them into NCCL. Figure~\ref{fig:ll-apis} provides an overview. Due to space constraints, we describe only the key primitives below. Full documentation and additional examples are available in the released source code\footnote{\url{https://github.com/ss16118/low-latency-nccl}}.

\paragraph{\texttt{ncclLLBuffer}}

\begin{figure}[!htp]
  \centering
  \includegraphics[width=1\linewidth]{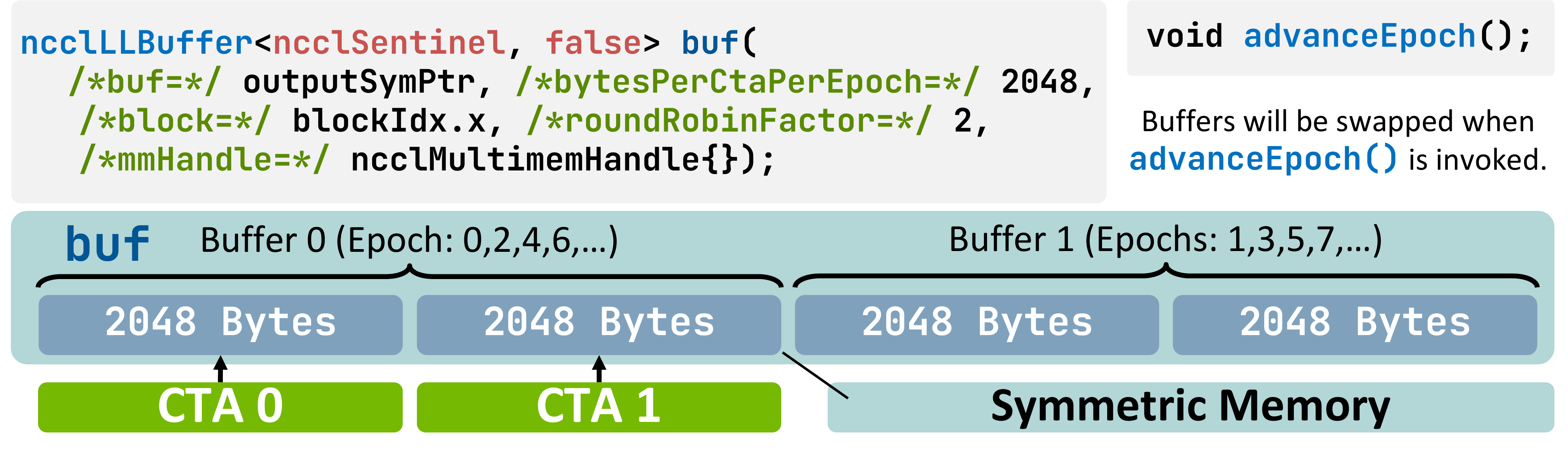}
  \cprotect\caption{Example construction and layout of a \texttt{ncclLLBuffer} object. Each CTA is assigned a fixed-size region per epoch, while epochs alternate between sub-buffers. Invoking \texttt{advanceEpoch()} switches the active sub-buffer and increments the internal \texttt{epoch} value.}
  \label{fig:ncclLLBuffer-constructor}
\end{figure}

To improve flexibility, we adopt a \textbf{buffer-centric} design for the new APIs, centered on the device-side \texttt{ncclLLBuffer} object. This abstraction wraps arbitrary symmetric memory and exposes low-latency primitives that operate directly on the buffer. A unified interface supports both LL and sentinel modes, selectable via the \texttt{ncclLLSyncMode} template parameter at initialization, allowing users to switch synchronization mechanisms with minimal changes. The LL128 method is not included because it requires 8 threads to operate as a unit, which does not fit the thread-level API design and is incompatible with the proposed primitives. A second template parameter controls whether NVLS multicast instructions are used.

To support multi-buffered execution, \texttt{ncclLLBuffer} organizes memory according to parameters such as \texttt{bytesPerCtaPerEpoch} and \texttt{roundRobinFactor}. Buffer addresses are derived from the current epoch and CTA index, allowing different iterations to operate on disjoint memory regions. Figure~\ref{fig:ncclLLBuffer-constructor} shows an example partitioning of a symmetric buffer under this scheme. When \texttt{roundRobinFactor} is set to 0, the buffer is no longer subdivided, the offset remains fixed, and \texttt{advanceEpoch()} performs no operation. In this mode, users need to manage buffer offsets explicitly. The object also maintains an internal \texttt{epoch} value, which is used as the flag transmitted alongside the data when the LL protocol is selected.

\begin{figure}[!htp]
  \centering
  \includegraphics[width=1\linewidth]{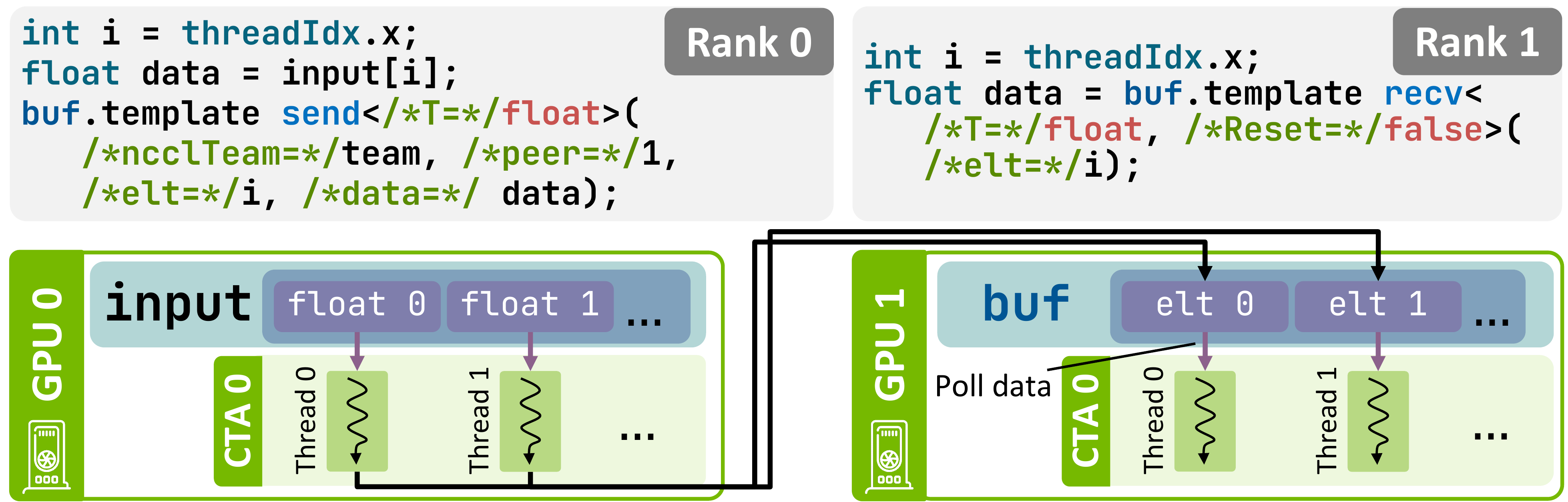}
  \cprotect\caption{Illustration of \texttt{send()} and \texttt{recv()} of a \texttt{ncclLLBuffer} for a single CTA.}
  \label{fig:ncclLLBuffer-send-recv}
\end{figure}

\paragraph{\texttt{send}}

The \texttt{send} primitive writes a value into a specified peer slot. The destination layout is determined by the type \texttt{T}: each element occupies \texttt{sizeof(T)} bytes, or \(2\times\texttt{sizeof(T)}\) for LL mode. This layout is used consistently across all API functions templated on \texttt{T}. In LL mode, the payload is packed with its flag whose size also equals \texttt{sizeof(T)}.
In sentinel mode, the receiving buffer must be initialized with sentinel values prior to transmission, and it is the user’s responsibility to ensure this.

\paragraph{\texttt{recv}}

The \texttt{recv} primitive polls a single buffer slot until valid data is observed, according to the chosen synchronization mode. In sentinel mode, it waits until the value differs from the sentinel, whereas in LL mode it waits until the flag matches the current epoch. The template parameter \texttt{T} used for \texttt{recv} must match that of the corresponding \texttt{send} operation. After reading, the slot can optionally be reset for reuse, as controlled by the \texttt{Reset} parameter. The visibility of the reset value is not guaranteed after the function returns. Users can enforce visibility by issuing a \texttt{\_\_threadfence()}.

\paragraph{\texttt{recvUnrolled}}

\texttt{recvUnrolled} extends \texttt{recv} to support receiving multiple elements with compile-time unrolling. It allows users to specify minimum and maximum element counts, enabling the compiler to generate optimized polling code. Elements in the range \texttt{[0, MinEltCount)} are always accessed, while elements in \texttt{[MinEltCount, MaxEltCount)} are accessed conditionally based on \texttt{eltCount}. This design is especially useful when receiving data from multiple peers and can significantly improve performance when \texttt{eltCount} is known in advance.

\paragraph{\texttt{recvReduce}}

The \texttt{recvReduce} primitive combines reception and reduction. It receives elements from multiple peers, converts them to an accumulator type, and applies a user-defined reduction operator, returning the accumulated result. Internally, it leverages \texttt{recvUnrolled()} for data reception and inherits its compile-time unrolling parameters.

\paragraph{\texttt{bcast}}

The \texttt{bcast} primitive writes a value to all peers simultaneously. When multicast support is enabled, the implementation uses hardware multicast to distribute data in a single operation. Otherwise, it iterates over peers. The template unroll factor controls loop expansion for improved throughput when broadcasting to multiple ranks.

\paragraph{\texttt{reset}}

The \texttt{reset} operation clears a specified buffer slot. In sentinel mode, it restores the sentinel value corresponding to the given type \texttt{T}, whereas in LL mode it sets the slot contents to zero. Resetting does not guarantee global visibility, and users may enforce it with an explicit memory fence. Although functions such as \texttt{recv()} and \texttt{recvReduce()} can optionally perform a reset after data reception, providing a dedicated primitive offers finer control over the buffer and allows users to decouple it from data reception when needed. The \texttt{resetRange()} function supports resetting multiple slots, but is omitted here for brevity.

\paragraph{Host-side Functions}

The API also provides host-side utilities. The function \texttt{ncclCalcScratchBufferSize()} computes the minimum buffer size needed for a given configuration, accounting for element size, rank count, CTA count, synchronization mode, and the round-robin factor, which determines the number of sub-buffers required for double or multiple buffering. The initialization function \texttt{ncclLLBufferInitSentinel()} prepares the given buffers for sentinel mode by filling them with type-specific sentinel values.

The host interface is intentionally minimal to simplify the development of custom kernels. Users only need to allocate and initialize symmetric buffers as usual, without additional orchestration. Once wrapped by the device-side abstraction, all necessary low-latency primitives will be exposed, allowing developers to focus on algorithm design rather than the setup.

\paragraph{Constraints}

The low latency APIs presented here are intentionally fine grained. Unlike NVSHMEM, which provides APIs at the thread, warp, and block levels, our design exposes only thread level primitives. This choice maximizes user control and allows implementations to achieve the lowest possible latency. Moreover, the APIs favor bidirectional communication patterns to achieve safety and optimal performance. As an example, algorithms with independent per-iteration communication, such as one-shot and two-shot AllReduce, can operate without barriers because each step uses disjoint buffers and each rank both sends and receives within the same iteration. In contrast, algorithms with inter-step dependencies and without bidirectional communication, such as ring AllReduce, still require explicit memory barriers. In a ring schedule, each rank must receive a chunk before reducing and forwarding it, as advancing early risks overwriting unconsumed data.

\paragraph{NCCL Collective Integration}
\label{nccl-collective-api-integration}
Using the proposed APIs, we implement one-shot, two-shot, and two-shot LL128 atomic AllReduce, along with other collectives. Taking AllReduce as an example, algorithms can be selected via \texttt{NCCL\_SYM\_KERNEL}: \texttt{AllReduce\_LLBuffer} for one-shot, \texttt{AllReduce\_LLBuffer\_Twoshot} for two-shot, and \texttt{AllReduce\_LL128\_Atomic} for the LL128 atomic variant. The synchronization mode is controlled by \texttt{NCCL\_SYM\_LLBUFFER\_SYNC}, allowing users to switch between LL and sentinel mechanisms for the same algorithm. Additionally, two-shot AllReduce requires symmetric (LSA) output buffers, while one-shot only relies on symmetric scratch buffers initialized by NCCL and has no such requirements.

The released artifact also includes low-latency Broadcast, Reduce, ReduceScatter, and AllGather kernels constructed from the same LL and sentinel primitives.

\subsection{Example One-shot AllReduce}

\begin{figure}[!htp]
  \centering
  \includegraphics[width=1\linewidth]{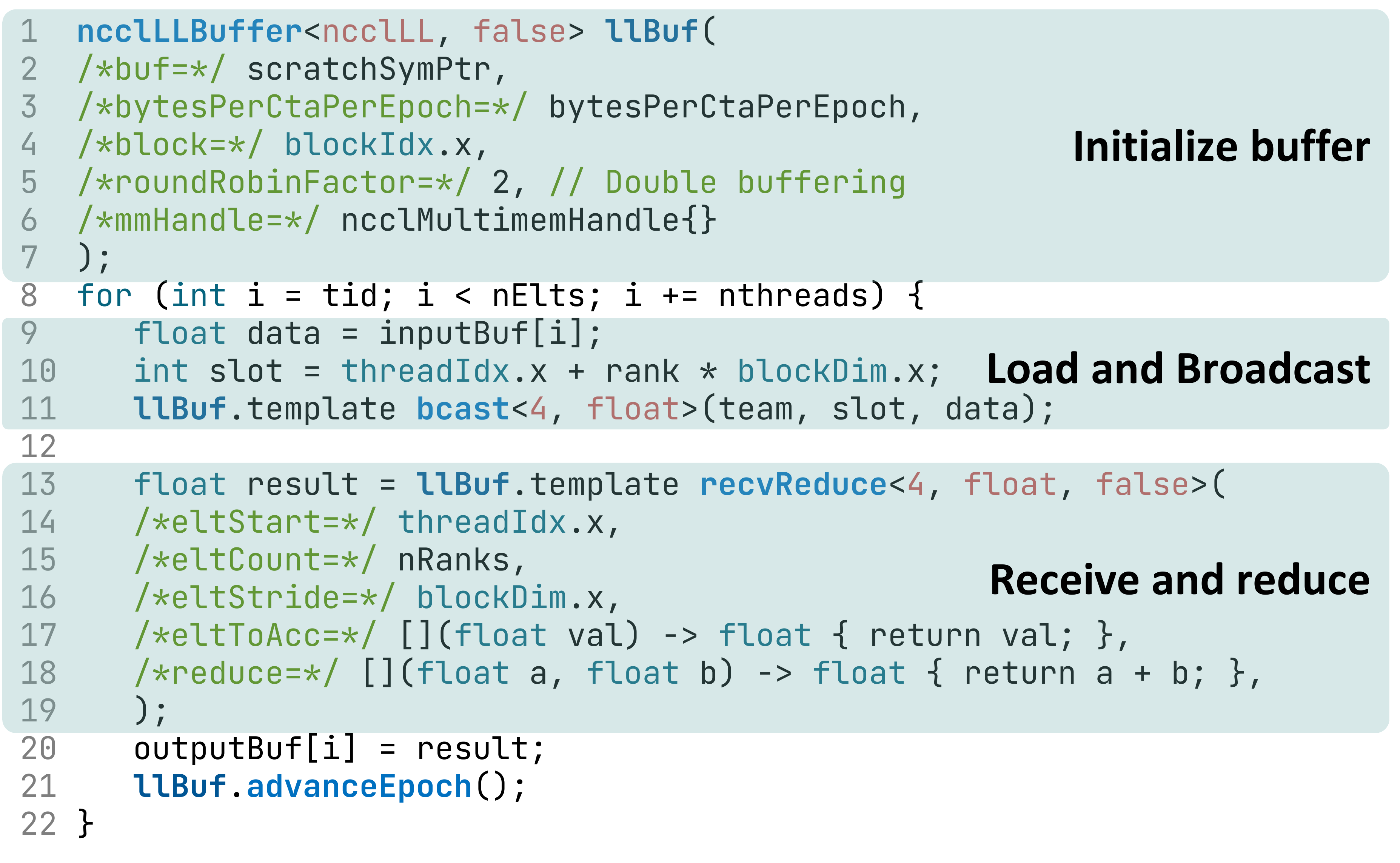}
  \cprotect\caption{Example implementation of a one-shot AllReduce using the low-latency API with \texttt{ncclLL} synchronization. Colored regions highlight the three stages of the algorithm.}
  \label{fig:one-shot-ar-api-example}
\end{figure}

Fig.~\ref{fig:one-shot-ar-api-example} shows a one-shot AllReduce implemented with the proposed low-latency API that uses \texttt{ncclLL} synchronization and supports arbitrary message sizes. A \texttt{ncclLLBuffer} is first constructed over symmetric scratch memory with double buffering. Each thread iterates over its assigned elements, loads a value from the input buffer, and invokes \texttt{bcast} to distribute it to peer slots. It then calls \texttt{recvReduce} to poll, receive, and reduce values from all ranks. The result is written to the output buffer, and \texttt{advanceEpoch()} switches to the next sub-buffer. This example illustrates that the API expresses collective algorithms succinctly while reducing the effort required to implement custom low-latency kernels.

\section{Measuring the Speed-of-Light of AllReduce}

\begin{figure}[!htp]
  \centering
  \includegraphics[width=1\linewidth]{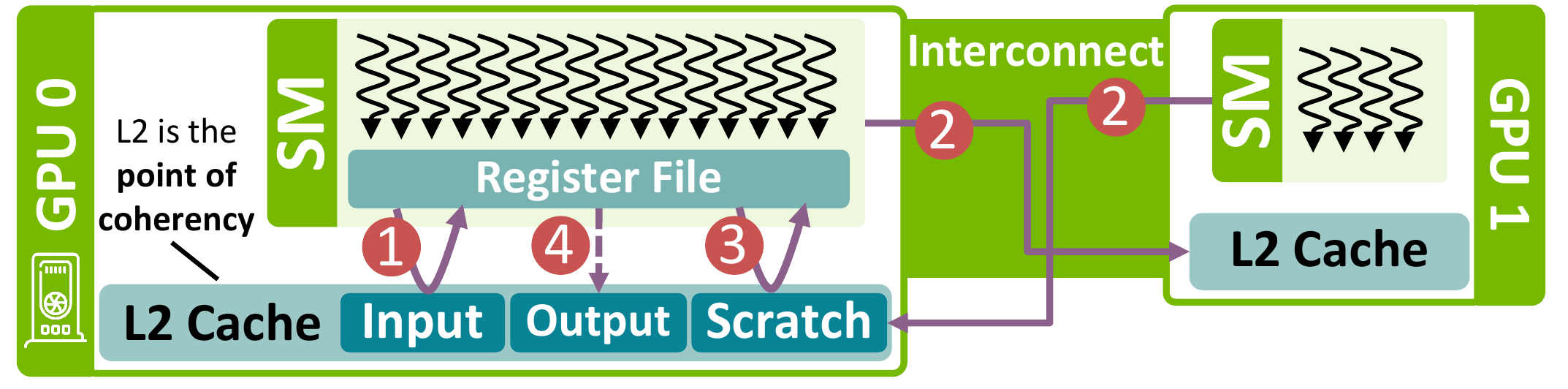}
  \cprotect\caption{Minimal data movement in an AllReduce. All buffers are assumed to reside in L2, which serves as the point of coherency (PoC) across GPUs in NVIDIA's systems. \textbf{SM} (streaming multiprocessor) is used instead of CTA to emphasize the hardware unit executing the memory operations.}
  \label{fig:sol-ar}
\end{figure}

In this section, we describe how the SoL lower bound can be estimated. We define the SoL as the minimal data movement required to complete the AllReduce while ignoring \textbf{all other overheads}, such as instruction scheduling and computation. Since small-message latency is determined by the transfer of the smallest unit handled by the memory system, we focus on the movement of a single 128-byte cache line.

Fig.~\ref{fig:sol-ar} illustrates the minimal data movements required for an AllReduce operation. Since remote stores are faster than remote loads and sufficient buffer space is assumed, the SoL bound corresponds to a one-shot push algorithm. The resulting data movement consists of the following components.

\noindent\circled{1}\hspace{0.3em}
The first step loads the data into the SM register file. To obtain the lower bound, we assume an L2 cache hit, which is reasonable since the message sizes considered typically fit in L2. This load incurs one L2 RTT, denoted as $L_{\text{L2\_RTT}}$.

\noindent\circled{2}\hspace{0.3em}
The data is then broadcast to the scratch buffers of all GPUs. For the SoL estimate, we assume that the stores to all peers are issued simultaneously. The data becomes visible in the remote GPU L2 cache after a latency of $L_{\text{remote\_store}}$. The cost of writing to the local scratch buffer is ignored because it is much smaller than the latency of remote stores.

\noindent\circled{3}\hspace{0.3em}
Once the data arrives, it is immediately loaded from L2 into the SM. Assuming an L2 hit, this incurs another $L_{\text{L2\_RTT}}$. Contributions from all peers are assumed to arrive simultaneously and are fetched in parallel.

\noindent\circled{4}\hspace{0.3em}
Reduction is performed inside the SM, and the result is written to the output buffer. The latency of this final store is not included because once the store instruction is issued, the memory system will finish the write in the background.

Combining the components above, the SoL latency of an AllReduce can be expressed as
\[
  L_{\text{SoL}} = 2L_{\text{L2\_RTT}} + L_{\text{remote\_store}}.
\]
Under the SoL assumption, sending data to $N$ peers incurs the same latency as sending to a single peer. Therefore, \textbf{$L_{\text{SoL}}$ represents an absolute lower bound that is independent of the number of ranks involved}.

To measure $L_{\text{L2\_RTT}}$, we benchmark the latency of a single \texttt{\_\_threadfence()}. This instruction enforces ordering and visibility at the L2 level, forcing the SM to wait until outstanding memory transactions reach the cache. Hence, its latency provides a good approximation of the L2 RTT.

To estimate $L_{\text{remote\_store}}$, we measure the RTT when a single value is ping-ponged between two GPUs, which can be decomposed as
$L_{\text{ping\_pong}} = 2L_{\text{L2\_RTT}} + 2L_{\text{remote\_store}}$. Each round-trip involves one remote store to the peer GPU and one remote store in the return direction, with an L2 access on both sides. Therefore, we can estimate $L_{\text{remote\_store}}$ as
\[
  L_{\text{remote\_store}} = (L_{\text{ping\_pong}} - 2L_{\text{L2\_RTT}}) / 2 .
\]
On two GB200, we measured the $L_{\text{L2\_RTT}}$ and $L_{\text{remote\_store}}$ as 0.306~µs and 0.792~µs, respectively. Therefore, the SoL latency of an AllReduce is computed to be \textbf{1.404~µs}.

\begin{figure*}[!htp]
  \centering
  \includegraphics[width=1\linewidth]{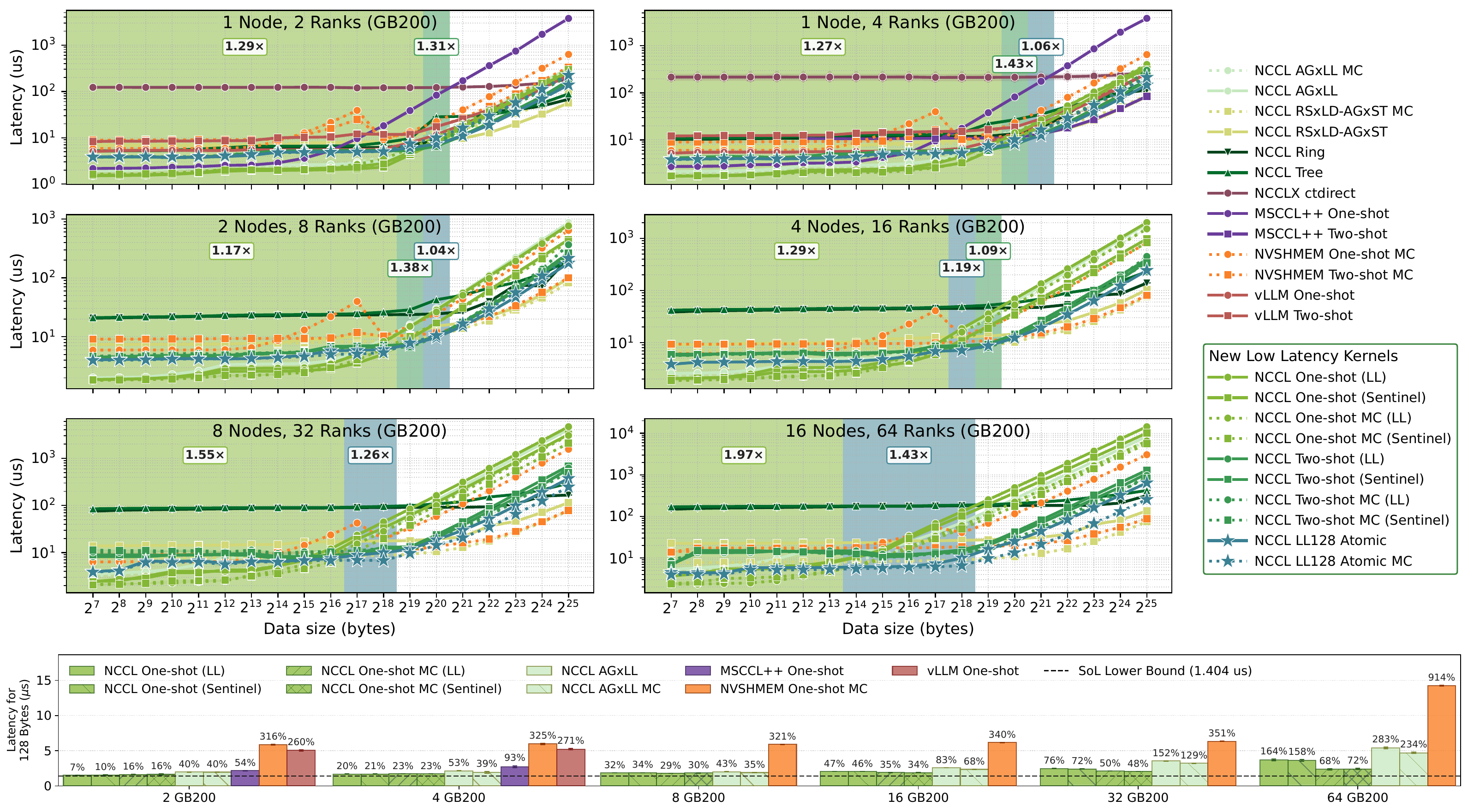}
    \cprotect\caption{Top plot shows out-of-place AllReduce latency versus message size on GB200 for 2 to 64 GPUs. Each subplot shows the mean over 10 trials for AllReduce implementations from NCCL, NCCLX, NVSHMEM, MSCCL++, and vLLM. Shaded regions mark message sizes where one of our kernels is fastest, regardless of the synchronization mode: \textcolor[HTML]{85B737}{light green} for \texttt{LLBuffer} one-shot, \textcolor[HTML]{399952}{dark green} for \texttt{LLBuffer} two-shot, and \textcolor[HTML]{3A8193}{blue-gray} for LL128 atomic. Labels within these regions report the geometric-mean speedup over the fastest existing implementation at each message size in the region. Dashed lines indicate multicast variants. Bottom plot shows the latency at 128B for selected one-shot kernels across GPU counts. The dashed horizontal line marks the measured SoL lower bound. Percentages report the overhead of each kernel relative to this bound.}
  \label{fig:latency-microbenchmarks}
\end{figure*}

\section{Microbenchmarks}
\label{sec:microbenchmarks}

\subsection{Experimental Setup}

We evaluate microbenchmarks on a GB200 NVL72 system, where 4 Blackwell GPUs reside within a node and 72 GPUs are connected within a single NVLink domain with 130~TB/s aggregate bandwidth. To ensure reproducibility, we use the NVIDIA vLLM container (v26.02)~\cite{nvidia2026vllmcontainer}, which includes Ubuntu 24.04, CUDA 13.1, vLLM 0.15.1, PyTorch 2.11, and OpenMPI 4.1.9. Each experiment is repeated over 10 trials, and we report the mean. Error bars denote standard deviation and are typically too small to be visible.

\subsection{Impact of Scratch Buffer}

\begin{figure}[!htp]
  \centering
  \includegraphics[width=1\linewidth]{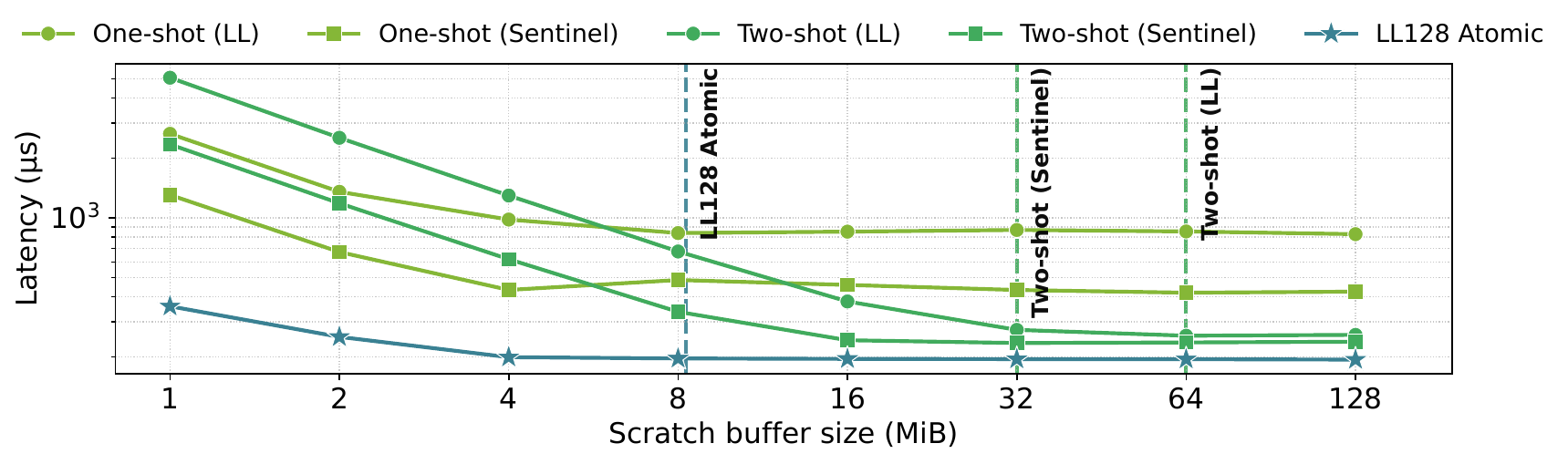}
  \cprotect\caption{Impact of scratch buffer size on AllReduce latency for our \texttt{LLBuffer}-based kernels on 2 nodes with 8 GB200. Vertical dashed lines indicate the minimum buffer size required for each kernel to process the full message in a single iteration.}
  \label{fig:scratch-buffer-sweep}
\end{figure}

Before benchmarking AllReduce latency, we first study the impact of scratch buffer capacity to determine practical default sizes for our kernels. Fig.~\ref{fig:scratch-buffer-sweep} shows results for \texttt{LLBuffer}-based kernels on 8 GPUs with a 32~MiB message. All kernels use 64 CTAs with up to 512 threads per CTA.

When the buffer is small ($<$ 8~MiB), two-shot kernels are slower than one-shot kernels. Each iteration of the two-shot design requires two synchronization steps, which dominate latency. Although fewer iterations are needed overall, the additional synchronization cost outweighs this benefit. As the buffer size increases, the performance of two-shot kernels improves rapidly and then plateaus once the entire message fits within a single iteration. This indicates that, ideally, achieving the best performance for two-shot kernels requires a scratch buffer large enough to process the data in one iteration.

In one-shot kernels, every rank writes to all peers simultaneously, which means that increasing the buffer size increases the amount of data sent per iteration, which raises instantaneous bandwidth pressure on the NVLink fabric. Hence, beyond a certain point, increasing the scratch size provides little benefit and can even slightly degrade performance.

The LL128 atomic kernel is particularly attractive because it is both space efficient and fast. Instead of storing intermediate data from all ranks, it accumulates contributions directly into the destination buffer, significantly reducing the scratch space needed to process the entire message.

Based on these results, we select 4~MiB of scratch buffer for one-shot kernels, as they show limited benefit from larger buffers. For two-shot and LL128 atomic kernels, we choose 64~MiB as the default.

\subsection{Latency}

With the scratch buffer size fixed, we perform a comprehensive evaluation across various data sizes and GPU counts. We compare our new low-latency kernels against the implementations from the state-of-the-art libraries and frameworks, including NCCL's legacy ring, tree algorithms, and symmetric memory kernels (v2.29.1), NVSHMEM (v3.5.21), MSCCL++ (v0.8.0), and vLLM custom AllReduce (v0.15.1). For all NCCL kernels, we used up to 64 CTAs with 512 threads per CTA. For other libraries, we used their default configurations.

Results are shown in Fig.~\ref{fig:latency-microbenchmarks}. In the bottom plot, we zoom in on the case where the message size is a single cache line, and compare the latency of different one-shot AllReduce implementations to the SoL bound. We only show the data for 32-bit floats, as the latency for 16-bit floats (i.e., float16 and bfloat16) is very similar.

\begin{figure*}[!htp]
  \centering
  \includegraphics[width=1\linewidth]{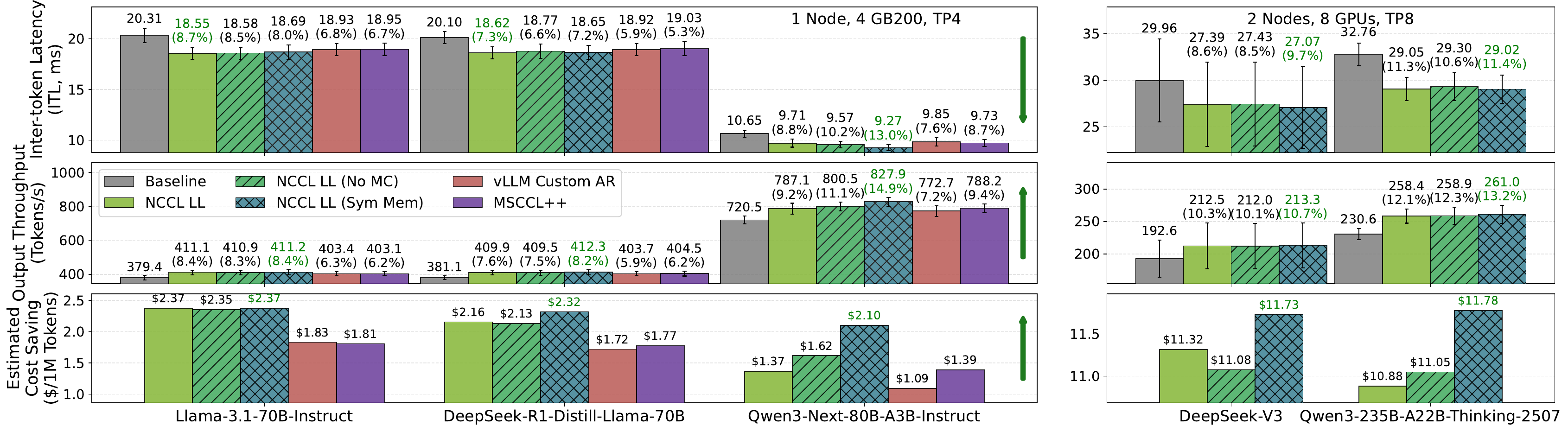}
  \cprotect\caption{Effect of low-latency collectives on vLLM inference. Rows report mean inter-token latency (ITL), output throughput, and estimated cost savings per 1M output tokens relative to the baseline. Percentages show the improvement over the baseline. Green labels highlight the best-performing configuration for each model. Green arrows indicate the direction of improvement.}
  \label{fig:vllm-inference-data}
\end{figure*}

For NCCL, the \texttt{AGxLL} and \texttt{RSxLD-AGxST} kernels correspond to its latest one-shot and two-shot symmetric AllReduce algorithms, respectively. For NCCLX CTran, we use the \texttt{ctdirect} algorithm, which supports only single-node execution, so results are reported for 2 and 4 GPUs. The ring variant (\texttt{ctring}) is omitted, as it supports only one rank per node and is not optimized for low latency. The same single-node limitation applies to vLLM’s custom AllReduce. Although MSCCL++ provides one-shot and two-shot AllReduce algorithms, including multicast variants, the multicast implementations consistently hang on GB200 and are excluded. The non-multicast variants are evaluated only on 2 and 4 GPUs, since multi-node collectives are not supported at the time of writing. A hierarchical AllReduce does exist for 2 nodes, but only with 8 GPUs per node. The two-shot algorithm also does not support 2 ranks or message sizes below 4~KiB, resulting in missing data points in the figure.

Based on the results, we make the following observations:

\paragraph{Observation 1}
Our \texttt{LLBuffer}-based one-shot AllReduce achieves the lowest latency for small messages across all GPU counts. As shown in Fig.~\ref{fig:latency-microbenchmarks}, our kernels incur only about 7\% overhead over the SoL bound at 2 GPUs, while competing implementations remain noticeably farther away. Even at 64 GPUs, multicast one-shot variants stay within roughly 70\% overhead of the SoL bound. While NCCL AGxLL and MSCCL++ also use LL-style synchronization, our implementation benefits from targeted optimizations including aggressive compile-time unrolling and parallel polling and reduction across ranks, which reduce detection and accumulation latency. We also observe a crossover between LL and sentinel synchronization. LL performs slightly better at very small sizes, whereas sentinel becomes preferable as message size and rank count increase, avoiding the flag overhead of LL.

\paragraph{Observation 2}
When the number of GPUs is small, the advantage of the LL128 atomic kernel over the standard two-shot design is limited. L2 atomic operations incur slightly higher latency due to serialization compared to accumulating values in the scratch buffer. As the number of GPUs increases, however, the scalability of atomic operations becomes more apparent, allowing the LL128 kernel to outperform two-shot kernels over a wider range of small and medium message sizes.

\paragraph{Observation 3}
At small scale, hardware multicast may incur slight overhead compared to unicast, but it plays a key role in improving collective scalability. The kernels that outperform our \texttt{LLBuffer}-based designs are primarily multicast variants that leverage the \texttt{multimem.ld\_reduce} instruction to combine contributions across ranks. This greatly reduces both the number of explicit memory operations and the amount of software-managed reduction. The benefit becomes more pronounced at larger scales, where offloading reduction to the NVLink/NVSwitch fabric is particularly effective~\cite{khalilov2024networkoffloaded}.

From the microbenchmarks, we observe that the proposed \texttt{LLBuffer}-based kernels cannot fully replace the existing symmetric kernel, namely the two-shot \texttt{RSxLD-AGxST}, as their polling overhead increases with GPU count and message size. However, they significantly reduce latency for small to medium messages, approaching the hardware limit, thereby complementing existing kernels.

\section{Case Studies}

After extensively benchmarking the kernels, we modify NCCL to select the best-performing low-latency algorithm based on the collected empirical results. For example, at 4 ranks, messages below 1 MiB use one-shot kernels, while those between 1 and 2 MiB use two-shot kernels. In this section, we evaluate the impact of these selections on two target workloads, LLM inference and cuSOLVERMp.

\subsection{LLM Inference}

For LLM inference, we conduct experiments on the same NVL72 GB200 system described earlier, using the same NVIDIA vLLM container from Section~\ref{sec:microbenchmarks}. We select vLLM as the inference framework due to its wide adoption and active development~\cite{kwon2023vllm, kolluru2025comparativeanalysis, neosignal_vllm_radar}. Experiments are conducted on several popular open-weight LLMs under two configurations: 1 node with 4 GPUs (TP=4) and 2 nodes with 8 GPUs (TP=8).

For all models, we use long input contexts of 100–200k tokens and generate 16K output tokens, with a batch size of 8 to emulate long-context inference scenarios. This choice reflects the growing importance of long-context workloads in practice~\cite{denain2025longcontext, chung2025longcontext, zhang2024chainagentslargelanguage, sun2025breakingboundarieslongcontextllm}. Each setting is evaluated over 5 trials with vLLM's serving benchmark. We present the results in Fig.~\ref{fig:vllm-inference-data}.

We evaluate several configurations against the NCCL baseline using only legacy kernels. \textit{NCCL LL} employs the proposed \texttt{LLBuffer}-based one-shot kernels, while \textit{No MC} disables NVLS multicast. \textit{Sym Mem} enables PyTorch symmetric memory, registering input and output buffers as symmetric memory to unlock NCCL’s two-shot and LL128 atomic kernels. Without it, AllReduce is limited to one-shot LL kernels and falls back to legacy implementations beyond the LL threshold. We report vLLM custom AllReduce and MSCCL++ only for the single-node case, since both are restricted to single-node execution as mentioned previously.

Across all models, low-latency kernels consistently improve performance. The best configuration reduces ITL by 7--13\% on 4 GPUs and 9--11\% on 8 GPUs, with similar gains in throughput. Results hold across diverse model architectures, including dense (Llama), mixture-of-experts (DeepSeek), and hybrid attention (Qwen3-Next). \textit{NCCL LL} alone performs well because decode steps involve small AllReduce operations, where one-shot \texttt{LLBuffer} kernels approach the SoL bound. \textit{Sym Mem} provides additional gains by enabling more efficient two-shot kernels for larger messages, particularly in vLLM’s mixed prefill/decode execution where some operations exceed the LL threshold. The benefit is more pronounced in throughput than ITL, as throughput measures total generated tokens over wall-clock time and thus captures improvements across the entire generation process, including prefill.

We estimate cost savings by converting output throughput into dollars per 1M tokens using CoreWeave's GB200 pricing~\cite{coreweave2026pricing}. For hourly price $p$ and output throughput $r$, the estimated cost is $p \cdot 10^6 / (3600r)$. We recognize that production serving is often disaggregated~\cite{bentoml2025pddisaggregation, li2026sloaware}, with separate node pools for prefill and decode, so this is not a full deployment cost model. Rather, it is an estimate of the savings attributable to faster collectives. Under this estimate, gains exceed \$11 per 1M output tokens in the 8-GPU setting for large models such as DeepSeek V3. In the 4-GPU case, although per-token savings are smaller, they accumulate into meaningful cost reductions for decode-heavy workloads at production scale.

\subsection{Traditional HPC}

To evaluate a representative traditional HPC workload, we use cuSOLVERMp. Many production GPU-accelerated applications, including GROMACS and LAMMPS, still rely on MPI or CUDA-aware MPI rather than NCCL~\cite{kraus2025cudaawarempi,gromacs2024install,lammps2026gpu,myers2024amrex,huebl2025amrexnccl}. Although QMCPACK~\cite{kim2018qmcpack} exposes NCCL support, it is limited to the Auxiliary-Field Quantum Monte Carlo (AFQMC) method and is not well maintained~\cite{qmcpack2023issue4654}, so we exclude it. NVIDIA HPCG, the High Performance Conjugate Gradients benchmark adapted to use NCCL, is a potential alternative but is also excluded as it relies exclusively on point-to-point communication~\cite{nvidia_hpcg}. We therefore use cuSOLVERMp, NVIDIA’s distributed dense linear algebra library~\cite{nvidia2026cusolvermp}, as a practical case study. Its generalized symmetric-definite eigensolvers are widely used in electronic-structure workloads~\cite{marek2014elpa}.

\begin{figure}[!htp]
  \centering
  \includegraphics[width=.95\linewidth]{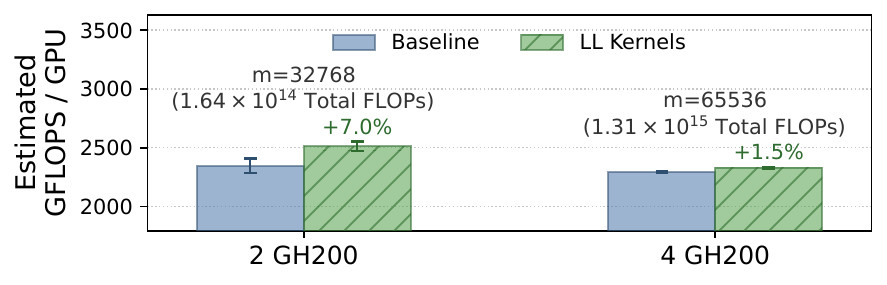}
  \cprotect\caption{Performance of \texttt{mp\_sygvd} with and without the new low-latency kernels. Bars show mean GFLOPS per GPU over 5 trials, with error bars indicating standard deviation. Percentage annotations show the improvement over the baseline.}
  \label{fig:cusolvermp-data}
\end{figure}

The experiments were conducted on the Alps supercomputing cluster. Each Alps node is equipped with four NVIDIA Grace Hopper Superchips (GH200) connected via 150~GB/s NVLink for intra-node communication~\cite{fusco2024understanding,cscs_alps}. As Alps does not provide an NVSwitch-based scale-up fabric across multiple nodes, we restrict this study to a single node. We chose this platform over GB200 as it reflects a more accessible system for domain scientists. Experiments were performed in NVIDIA's PyTorch container (v25.10) running Ubuntu 24.04, CUDA 12.6, OpenMPI 4.1.7, and cuSOLVERMp 0.7.2. In our setup, larger matrix sizes ($m{=}32768$ and $m{=}65536$) exceed single-GPU memory capacity and are therefore executed in distributed mode. We exclude MSCCL++ from this comparison due to MPI errors and compare only against the NCCL baseline using legacy kernels.

Figure~\ref{fig:cusolvermp-data} shows that the proposed low-latency kernels consistently improve cuSOLVERMp across the two configurations. The gains are more pronounced for $m{=}32768$, where communication constitutes a larger fraction of runtime. Note that cuSOLVERMp does not register buffers as symmetric memory, so only one-shot kernels were used for message sizes below 1~MiB. Overall, these results indicate that low-latency collectives benefit not only LLM inference but also traditional HPC workloads as NCCL-based communication becomes more widely adopted.

\section{Related Work and Discussion}

We are aware that several frameworks and communication libraries also provide low-latency AllReduce implementations, such as SGLang~\cite{zheng2024sglang} and FlashInfer~\cite{ye2025flashinfer}. These are not included in our microbenchmark comparison, as their designs largely resemble the custom AllReduce in vLLM, and therefore exhibit similar performance characteristics. In TensorRT-LLM~\cite{nvidia2023trtllm}, AllReduce variants based on sentinel-style synchronization are available, but they still rely on global barrier flags for coordination, which introduces additional overhead compared to our designs.

There also exist libraries such as DeepEP~\cite{zhao2025deepep} and NCCL EP~\cite{goldman2026ncclep} that provide low-latency primitives tailored for expert parallel workloads. These optimizations target a narrower class of communication patterns and do not generalize to other workloads. In contrast, our API is general-purpose and can be used to implement similar functionality when needed.

We do not compare against NIXL~\cite{ranadive2026nixl}, as it targets a different design space. NIXL is primarily a transport and orchestration layer for GPU communication, focusing on scheduling and integration across heterogeneous backends rather than optimizing the latency of collectives.

There are several directions for future work. First, kernel selection is currently driven by empirical measurements and could be improved with an accurate performance model. Developing such a model would require detailed knowledge of GPU architecture, including factors such as warp scheduling and instruction-level behavior, which is beyond the scope of this work. Second, the current API design focuses on thread-level primitives to maximize performance. Future extensions could provide warp- or block-level abstractions with fewer constraints and improve usability.

\section{Conclusion}

In this work, we studied how to approach the absolute hardware lower bound for scale-up GPU collectives within a single NVLink domain and identified key principles for low-latency design. Building on NCCL's device-side APIs, we introduced a new set of APIs that simplify the construction of custom low-latency collective kernels. Using these APIs, we implemented several new AllReduce kernels that substantially reduce latency for small and medium messages, bringing SoL overhead down to about 7\% in the best case and consistently outperforming state-of-the-art frameworks. In vLLM inference, the best configuration reduces ITL by up to 13\% on 4 GPUs and 11\% on 8 GPUs across diverse LLMs, with similar throughput gains and estimated savings of more than \$11 per million output tokens for a large model such as DeepSeek-V3. We also observe consistent single-node speedups for cuSOLVERMp on the Alps supercomputer. Overall, these results demonstrate that low-latency optimization delivers measurable benefits in both AI inference and traditional HPC applications, and that the proposed APIs provide a practical foundation for building latency-critical collective kernels.

\section{Acknowledgments}
The authors thank Andrei Ivanov for his invaluable assistance in collecting the experimental data. This work would not have been possible without his support. The research was conducted as part of the FastTrackAI project at the Singapore-ETH Centre, which was established collaboratively between ETH Zurich and the National Research Foundation, Singapore. This research is supported by the National Research Foundation, Singapore (NRF), and the Ministry of Digital Development and Information (MDDI) under the AI Visiting Professorship (Award No. AIVP-2025-005). This work also received funding from the European Research Council (Project PSAP, No. 101002047). The authors used ChatGPT-5.4~\cite{gpt54chat2026} to assist with light editing and proofreading. All content and ideas remain the original work of the authors.

\bibliographystyle{ieeetr}
\bibliography{references}

\end{document}